\documentclass[prc,twocolumn,superscriptaddress,showpacs,
	preprintnumbers,a4paper,nofootinbib]{revtex4}

\usepackage{graphicx,color}
\usepackage{dcolumn}
\usepackage{bm}
\graphicspath{{../Figures/}}

\begin{document}

\title{$^3$He($\alpha,\gamma$)$^7$Be cross section at low energies}

\author{Gy.~Gy\"urky}\affiliation{Institute of Nuclear Research (ATOMKI), Debrecen, Hungary}
\author{F.~Confortola}\affiliation{Universit\`a di Genova and INFN Sezione di Genova, Genova, Italy}
\author{H.~Costantini}\affiliation{Universit\`a di Genova and INFN Sezione di Genova, Genova, Italy}
\author{A.~Formicola}\affiliation{INFN, Laboratori Nazionali del Gran Sasso (LNGS), Assergi (AQ), Italy}
\author{D.~Bemmerer}
	\affiliation{Istituto Nazionale di Fisica Nucleare (INFN), Sezione di Padova, via Marzolo 8, 35131 Padova, Italy}\affiliation{Forschungszentrum Dresden-Rossendorf, Postfach 510119, 01314 Dresden, Germany}
\author{R.~Bonetti}\affiliation{Istituto di Fisica Generale Applicata, Universit\`a di Milano and INFN Sezione di Milano, Italy}
\author{C.~Broggini}
 \affiliation{Istituto Nazionale di Fisica Nucleare (INFN), Sezione di Padova, via Marzolo 8, 35131 Padova, Italy}
\author{P.~Corvisiero}\affiliation{Universit\`a di Genova and INFN Sezione di Genova, Genova, Italy}
\author{Z.~Elekes}\affiliation{Institute of Nuclear Research (ATOMKI), Debrecen, Hungary}
\author{Zs.~F\"ul\"op}\affiliation{Institute of Nuclear Research (ATOMKI), Debrecen, Hungary}
\author{G.~Gervino}\affiliation{Dipartimento di Fisica Sperimentale, Universit\`a di Torino and INFN Sezione di Torino, Torino, Italy}
\author{A.~Guglielmetti}\affiliation{Istituto di Fisica Generale Applicata, Universit\`a di Milano and INFN Sezione di Milano, Italy}
\author{C.~Gustavino}\affiliation{INFN, Laboratori Nazionali del Gran Sasso (LNGS), Assergi (AQ), Italy}
\author{G.~Imbriani}\affiliation{Dipartimento di Scienze Fisiche, Universit\`a di Napoli ''Federico II'', and INFN Sezione di Napoli, Napoli, Italy}
\author{M.~Junker}\affiliation{INFN, Laboratori Nazionali del Gran Sasso (LNGS), Assergi (AQ), Italy}
\author{M.~Laubenstein}\affiliation{INFN, Laboratori Nazionali del Gran Sasso (LNGS), Assergi (AQ), Italy}
\author{A.~Lemut}\affiliation{Universit\`a di Genova and INFN Sezione di Genova, Genova, Italy}
\author{B.~Limata}\affiliation{Dipartimento di Scienze Fisiche, Universit\`a di Napoli ''Federico II'', and INFN Sezione di Napoli, Napoli, Italy}
\author{V.~Lozza}\affiliation{Istituto Nazionale di Fisica Nucleare (INFN), Sezione di Padova, via Marzolo 8, 35131 Padova, Italy}
\author{M.~Marta}\affiliation{Istituto di Fisica Generale Applicata, Universit\`a di Milano and INFN Sezione di Milano, Italy}
\author{R.~Menegazzo}\affiliation{Istituto Nazionale di Fisica Nucleare (INFN), Sezione di Padova, via Marzolo 8, 35131 Padova, Italy}
\author{P.~Prati}\affiliation{Universit\`a di Genova and INFN Sezione di Genova, Genova, Italy}
\author{V.~Roca}\affiliation{Dipartimento di Scienze Fisiche, Universit\`a di Napoli ''Federico II'', and INFN Sezione di Napoli, Napoli, Italy}
\author{C.~Rolfs}\affiliation{Institut f\"ur Experimentalphysik III, Ruhr-Universit\"at Bochum, Bochum, Germany}
\author{C.~Rossi Alvarez}\affiliation{Istituto Nazionale di Fisica Nucleare (INFN), Sezione di Padova, via Marzolo 8, 35131 Padova, Italy}
\author{E.~Somorjai}\affiliation{Institute of Nuclear Research (ATOMKI), Debrecen, Hungary}
\author{O.~Straniero}\affiliation{Osservatorio Astronomico di Collurania, Teramo, and INFN Sezione di Napoli, Napoli, Italy}
\author{F.~Strieder}\affiliation{Institut f\"ur Experimentalphysik III, Ruhr-Universit\"at Bochum, Bochum, Germany}
\author{F.~Terrasi}\affiliation{Seconda Universit\`a di Napoli, Caserta, and INFN Sezione di Napoli, Napoli, Italy}
\author{H.P.~Trautvetter}\affiliation{Institut f\"ur Experimentalphysik III, Ruhr-Universit\"at Bochum, Bochum, Germany}

\collaboration{The LUNA Collaboration}\noaffiliation


\begin{abstract}
The flux of $^7$Be and $^8$B neutrinos from the Sun and the production of $^7$Li via primordial nucleosynthesis depend on the rate of the $^3$He($\alpha$,$\gamma$)$^7$Be reaction. 
In extension of a previous study showing cross section data at 127 - 167\,keV center of mass energy, the present work reports on a measurement of the $^3$He($\alpha$,$\gamma$)$^7$Be cross section at 106\,keV performed at Italy's Gran Sasso
underground laboratory by the activation method. This energy is closer to the solar Gamow energy than ever reached before. The result is $\sigma$ = 0.567$\pm$0.029$_{\rm stat}$$\pm$0.016$_{\rm syst}$ nbarn. The data are compared with previous activation studies at high energy, and a recommended $S(0)$ value for all $^3$He($\alpha$,$\gamma$)$^7$Be activation studies, including the present work, is given. 
\end{abstract}


\maketitle

\section{Introduction}

The $^3$He($\alpha,\gamma$)$^7$Be and $^3$He($^3$He,2p)$^4$He reactions compete in the proton--proton (p--p) chain of solar hydrogen burning. The ratio of their rates at the temperature of the solar center determines how much the $^7$Be and $^8$B branches of the p--p chain contribute to solar hydrogen burning. The $^3$He($^3$He,2p)$^4$He cross section being comparatively well known \cite{Bonetti99-PRL}, the predicted flux of solar neutrinos from $^7$Be and
$^8$B decay \cite{BS05} depends on the $^3$He($\alpha,\gamma$)$^7$Be cross section: The 9\,\% uncertainty in its extrapolation to the solar Gamow energy (23\,keV) obtained in a global analysis \cite{Adelberger98-RMP} contributes 8\,\% \cite{Bahcall04-PRL} to the uncertainty in the predicted fluxes for solar $^7$Be and $^8$B neutrinos, in both cases the major nuclear contribution to the total uncertainty. The flux of solar $^8$B neutrinos has been measured in the SNO and SuperKamiokande neutrino detectors \cite{Aharmim05-PRC,SuperKamiokande06-PRD}, with a total uncertainty as low as 3.5\,\% \cite{SuperKamiokande06-PRD}. The solar $^7$Be neutrino flux is planned to be measured in the Borexino and KamLAND neutrino detectors. 

The production of $^7$Li in big-bang nucleosynthesis (BBN) is also
highly sensitive to the $^3$He($\alpha,\gamma$)$^7$Be cross section in the
energy range $E$ $\approx$ 160--380\,keV \cite{Burles99-PRL}. A recent compilation for the purpose of BBN adopts 8\,\% uncertainty \cite{Descouvemont04-ADNDT} for the cross section. Based on the
baryon to photon ratio from observed anisotropies in the cosmic microwave
background \cite{Spergel03-ApJSS}, nucleosynthesis network calculations predict primordial $^7$Li abundances \cite{Coc04-ApJ} that are significantly higher than
observations of old stars \cite{Ryan00-ApJL,Bonifacio02-AA}.
Either a completely new interpretation of the stellar abundance data \cite[e.g.]{Korn06-Nature} or a dramatically lower $^3$He($\alpha$,$\gamma$)$^7$Be cross section at relevant energies may explain this discrepancy.

Since the cross section of $^3$He($\alpha$,$\gamma$)$^7$Be reaction is of the order of attobarn at $E$ = 23 keV, the cross section data from experiments carried out at higher energies are parameterized by the astrophysical S factor $S(E)$ defined as
$$ S(E) = \sigma(E) \cdot E \exp(2\pi\eta(E)) $$
where $2 \pi \eta(E) = 164.12 \cdot E^{-0.5}$ is the Sommerfeld parameter \cite{Rolfs88}, and $E$ the center of mass energy in keV. The S factor is then used to extrapolate the data to the low energies of astrophysical interest, and often its extrapolation to zero energy, $S(0)$, is quoted. 

\begin{figure}[tb]
 \centering
  \includegraphics[angle=-90,width=0.9\columnwidth]{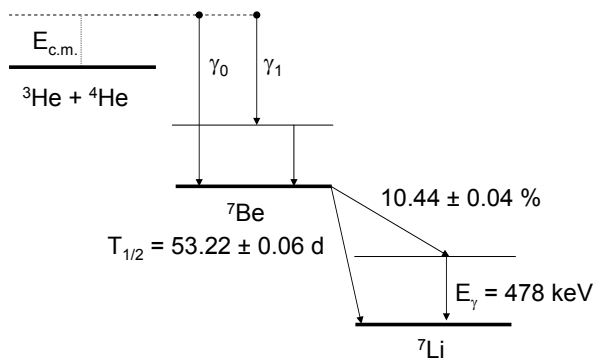}
 \caption{\label{fig:decay} Level diagram of the $^3$He($\alpha$,$\gamma$)$^7$Be reaction ($Q$ = 1.586\,MeV) and the decay of $^7$Be.
  } 
\end{figure}

The $^3$He($\alpha$,$\gamma$)$^7$Be reaction has a $Q$ value of 1.586\,MeV \cite{Audi03-AME}, and at low energy it proceeds via radiative capture into the ground state and the first excited state of $^7$Be (Fig.~\ref{fig:decay}). The final $^7$Be nucleus decays
with a half-life of 53.22\,$\pm$\,0.06\,days to $^7$Li, emitting a 478\,keV $\gamma$-ray
in 10.44\,$\pm$\,0.04\,\% of the cases \cite{Tilley02-NPA}. The cross section
can be measured by detecting either the induced $^7$Be activity (activation
method) or the prompt $\gamma$-rays from the reaction (prompt-$\gamma$ method)\footnote{An experiment based on a third method to measure the cross section, namely the detection of $^7$Be nuclei in a recoil mass separator, is in progress at the ERNA facility \cite{diLeva06-Kosproc}.}. 

Previous studies of the $^3$He($\alpha$,$\gamma$)$^7$Be reaction
\cite{Osborne82-PRL,Robertson83-PRC,Volk83-ZPA,NaraSingh04-PRL} that used the activation technique cover the
energy range $E$ = 420--2000\,keV and are briefly recalled here. 

Osborne {\it et al.} \cite{Osborne82-PRL} have measured the cross section by the activation technique at two energies, $E$ = 945 and 1250~keV. A $^3$He gas cell closed by a Ni+Cu window has been bombarded by an $\alpha$-beam, and the activity of $^7$Be implanted into a Ta catcher foil has been measured with a Ge(Li) detector.

A similar experimental technique (gas cell with Ni window, Au catcher foil, Ge(Li) detector) has been used by Robertson {\it et al.} \cite{Robertson83-PRC}. The beam intensity was measured by current integration as well as by Rutherford Backscattering (RBS) from a gold foil. An attempt was made to study the loss of $^7$Be from the catcher, giving a 20\,\% upper limit. The cross section was determined at $E$ = 987~keV, with both direct ($^3$He gas cell and $\alpha$-beam) and inverse ($^4$He gas cell and $^3$He-beam) kinematics yielding consistent results. 

Volk {\it et al.} \cite{Volk83-ZPA} measured the energy integrated cross section using a 0.8\,bar $^3$He gas cell in which the $\alpha$-beam stopped. The created $^7$Be was collected onto an Al foil. The energy dependence of the cross section from a previous prompt-$\gamma$ study was adopted in order to derive an $S(0)$ value.

Recently, Nara Singh {\it et al.} \cite{NaraSingh04-PRL} carried out a precise activation experiment at $E$ = 420 -- 950\,keV. A $^3$He gas cell closed with a Ni window has been bombarded with $\alpha$-beam. The beam intensity was measured by both current integration and RBS. The produced $^7$Be was collected on a Cu catcher and the activity was measured by a HPGe detector. 

Cross section measurements by the prompt $\gamma$-ray method
\cite{Holmgren59-PR,Parker63-PR,Nagatani69-NPA,Kraewinkel82-ZPA,Osborne82-PRL,Alexander84-NPA,Hilgemeier88-ZPA}
cover the energy range $E$ = 107--2500\,keV, although with limited precision at low energies. 
A global analysis of all available experimental data \cite{Adelberger98-RMP} indicates that S factor data
obtained with the activation method are systematically 13\,\% higher than
the prompt-$\gamma$ results. 

Theoretical calculations reproduce the global shape of the S factor curve rather well \cite[e.g.]{Kajino86-NPA,Csoto00-FBS,Marcucci06-NPA}. However,
the slope of this curve has been questioned \cite{Csoto00-FBS} for $E$ $\leq$
300\,keV, where there are no high-precision data. 

The aim of the present activation study is to provide high precision data at energies that are low enough to effectively constrain the extrapolation to solar energies and high enough to be relevant for big-bang nucleosynthesis. 
In order to study the solar interior
\cite{Bahcall04-PRL,Fiorentini03-arxiv,Bahcall05-astroph}, to investigate the low-energy slope of the S factor curve \cite{Csoto00-FBS} and to sharpen big bang $^7$Li abundance predictions \cite{Burles99-PRL,Serpico04-JCAP}, such precision $^3$He($\alpha$,$\gamma$)$^7$Be measurements have been recommended. 
In the present work, a new experimental cross section number is reported at $E$ = 106\,keV, lower than ever before reached by direct experiment. In addition, cross section data at $E$ = 127--169\,keV that have been published previously in abbreviated form \cite{Bemmerer06-PRL} are presented with full detail here. The impact of the present result for big-bang nucleosynthesis is analyzed, and a new $S(0)$ for the activation method based on all available experimental data is recommended.

\section{Experiment}

\begin{figure*}[tb]
 \centering
  \includegraphics[angle=-90,width=1.0\textwidth]{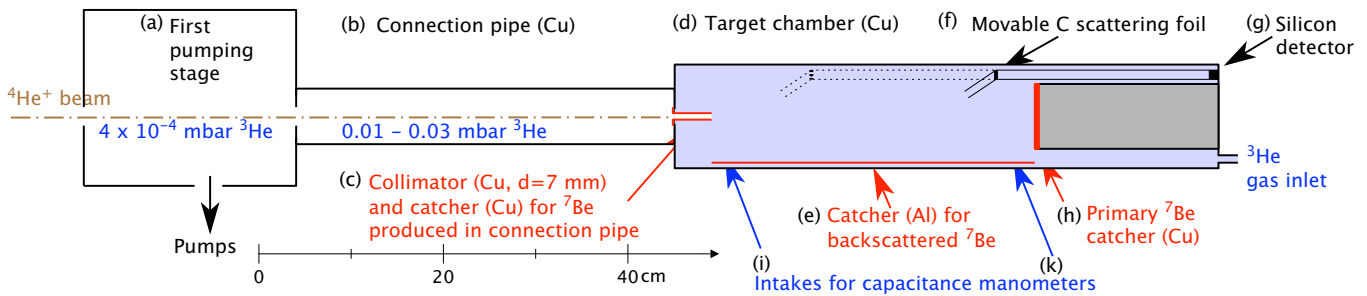}
 \caption{\label{fig:setup} (Color online) Schematic view of the target chamber used for the irradiations. See text for details.
  } 
\end{figure*}

\subsection{The accelerator}

The Laboratory for Underground Nuclear Astrophysics (LUNA) \cite{Greife94-NIMA} in Italy's Gran Sasso underground laboratory (LNGS) has been designed for measuring low nuclear cross sections for astrophysical purposes. Its low laboratory background \cite{Bemmerer05-EPJA} has made it possible to study several reactions of astrophysical relevance \cite{Bonetti99-PRL,Casella02-NPA,Formicola04-PLB,Imbriani05-EPJA,Lemut06-PLB}. 

The irradiations for the present study have been carried out at the 400\,kV LUNA2 accelerator \cite{Formicola03-NIMA} at energies $E_\alpha$ = 250, 300, 350 and 400\,keV, with a typical current of 200~$\mu$A $^4$He$^+$. The beam energy is obtained with an uncertainty as low as 300\,eV from a precision resistor chain calibrated through radiative-capture reactions, and it exhibits an energy spread of less than 100\,eV \cite{Formicola03-NIMA}.

The beam intensity is measured using a beam calorimeter with constant temperature gradient similar to the one described previously \cite{Casella02-NIMA}, and a precision of 1.5\,\% is obtained from the difference between the calorimeter power values with and without incident ion beam, taking into account the calculated energy loss in the target gas \cite{SRIM03-26}. The calorimeter has been calibrated at various beam energy and intensity values using the evacuated gas target chamber as a Faraday cup, with a proper secondary electron suppression voltage applied.

\subsection{The gas target setup}

The $^3$He($\alpha,\gamma$)$^7$Be reaction takes place in a differentially pumped windowless gas target (Fig.~\ref{fig:setup}) filled with enriched $^3$He gas (isotopic purity $>$99.95\,\%, pressure 0.7\,mbar, corresponding target thickness 8--10\,keV). The exhaust from the first pumping stage (2050 $\rm\frac{m^3}{h}$ Roots pump) and the second pumping stage (three 1000 $\rm\frac{l}{s}$ turbomolecular pumps) is compressed by a 500 $\rm\frac{m^3}{h}$ Roots blower and an oil-free forepump, cleaned in a getter-based gas purifier and recirculated into the target. After passing the three pumping stages (the one closest to the target is shown in Fig.~\ref{fig:setup}\,a) and a connection pipe (b), the ion beam from the accelerator enters the target chamber (d) through an aperture of 7\,mm diameter (c) and is finally stopped on a disk (h) of 70\,mm diameter that serves as the primary catcher for the produced $^7$Be and as the hot side of the beam calorimeter described above.

\begin{figure}[b!]
 \centering
  \includegraphics[width=1.0\columnwidth]{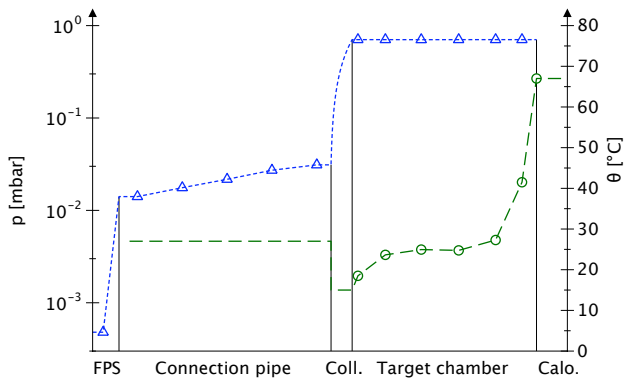}
 \caption{\label{fig:PressTempProfile} (Color online) Measured pressure ($p$, blue triangles) and temperature ($\theta$, green circles) profile inside the target chamber and adjacent regions: First pumping stage (FPS), connection pipe, collimator (Coll.), target chamber, calorimeter (Calo.). The dashed lines indicate the interpolated profile adopted where there are no data. 
  } 
\end{figure}

\begin{figure}[b!]
 \centering
  \includegraphics[width=1.0\columnwidth]{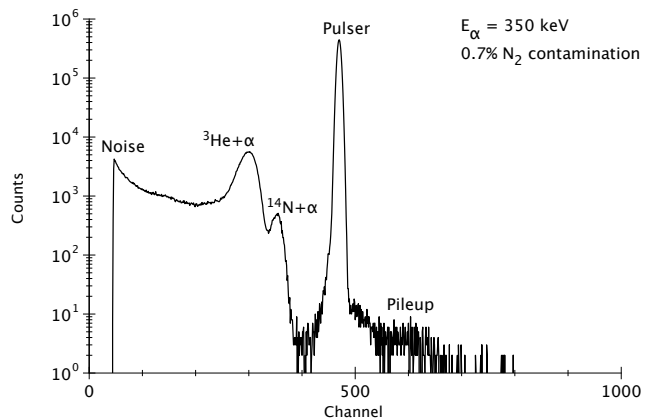}
 \caption{\label{fig:silicon350-051114} Elastic scattering spectrum taken with the silicon detector at $E_\alpha$ = 350\,keV, showing a contamination of 0.7\,\% N$_2$ in the $^3$He gas. 
  } 
\end{figure}

The pressure in the gas target chamber has been monitored continuously during the irradiations at two positions with capacitance manometers, (Fig.~\ref{fig:setup}\,i-k). The pressure and temperature profiles (Fig.~\ref{fig:PressTempProfile}) have been measured without ion beam in a chamber of the same dimensions as the actual gas target chamber but with several ports along the beam path for pressure and temperature sensors. The pressure has been found to be equal to better than 0.25\,\% at the different positions. The temperature profile has been observed to vary monotonously between the watercooled collimator (15\,$^\circ$C) and the hot side of the calorimeter (67\,$^\circ$C). Linear interpolations have been used to calculate pressure and temperature between the measured positions. 
In order to reflect the uncertainty from the linear interpolation, a relative uncertainty of 13\,\% has been assigned to the part of the target thickness contained in the 7\,mm collimator (which comprises 5\,\% of the total target thickness and where the pressure drop is significant), resulting in 0.7\,\% uncertainty for the total target thickness. 
%
Combining this uncertainty with the 0.25\,\% manometer precision and with the 0.3\,\% uncertainty from the temperature measurement, a precision of 0.8\,\% for the target thickness without ion beam is obtained. 

The thinning of the target gas through the beam heating effect \cite{Goerres80-NIM} and the fraction of gases other than $^3$He have been measured in order to obtain the effective target thickness. For this purpose, a 100\,$\mu$m thick silicon detector (Fig.~\ref{fig:setup}\,g) detects projectiles that have been elastically scattered first in the target gas and subsequently in a movable 15\,$\mu$g/cm$^2$ carbon foil (f). The beam heating effect has thus been investigated at several positions along the chamber in a wide beam energy and intensity range, and the average corrections shown in Table~\ref{tab:samples} were found. 
The amount of contaminant gases (mainly nitrogen) is monitored with the silicon detector during the irradiations (Fig.~\ref{fig:silicon350-051114}), kept below 1.0$\pm$0.1\,\% and corrected for in the analysis. Further details of the elastic scattering measurements are described elsewhere \cite{Marta05-Master,Marta06-NIMA}.

\subsection{Sample irradiation}

The catchers are irradiated with charges of 60--200\,C, accumulating $^7$Be activities of 0.03--0.6\,Bq. Table~\ref{tab:irradiations} shows details of the irradiations.

Calculations for the straggling of the $^4$He beam and of the produced $^7$Be nuclei in the $^3$He gas and for the emission cone of $^7$Be (opening angle 1.8--2.1\,$^\circ$) have been carried out and show that 99.8\,\% of the $^7$Be produced inside the target chamber, including the 7\,mm collimator, reaches the primary catcher.

\begin{table}[b]
\caption{\label{tab:irradiations} Details of the irradiations. In all cases the target pressure was 0.7\,mbar.}
\begin{ruledtabular}
\setlength{\extrarowheight}{0.1cm}
\begin{tabular}{lccrrc}
Sample & $E_\alpha$  & Target & \multicolumn{1}{c}{Irradiation}   & \multicolumn{1}{c}{Charge} & Average      \\[-1mm]
       &    [keV]    & gas    & \multicolumn{1}{c}{[days]} & \multicolumn{1}{c}{[Coulombs]}      & current [$\mu$A]   
         \\ \hline
D & 249.8 & $^3$He & 6.5 & 83 & 149 \\
B & 298.8 & $^3$He & 10.5 & 215 & 237 \\
A & 348.4 & $^3$He & 9.5 & 203 & 248 \\
F & 398.2 & $^3$He & 2.9 & 63 & 250 \\
E & 400.2 & $^4$He & 6.5 & 104 & 187 \\
\end{tabular}
\end{ruledtabular}
\end{table}

\subsection{Offline $^7$Be counting}

\begin{figure}[t]
 \centering
  \includegraphics[angle=-90,width=1.0\columnwidth]{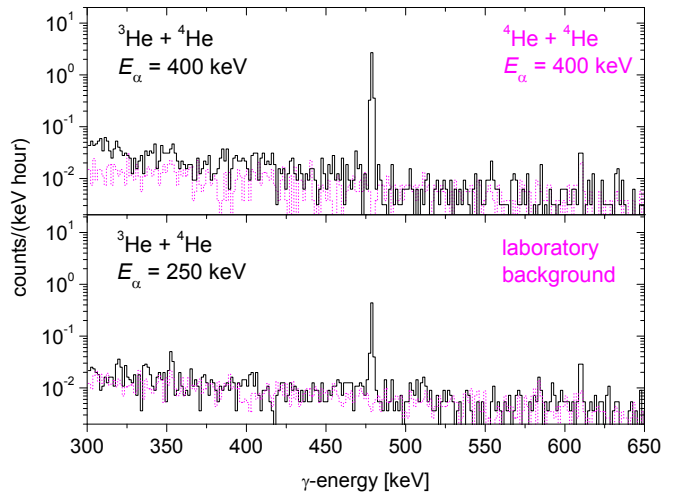}
 \caption{\label{fig:spectrum} 
(Color online) Offline $\gamma$-counting spectra, detector LNGS1. Solid black line: $^3$He gas bombarded at $E_\alpha$ = 400\,keV (top panel, sample F) and 250\,keV (bottom panel, sample D), respectively. Dotted red line, top panel: $^4$He gas bombarded at $E_\alpha$ = 400\,keV (sample E). Dotted red line, bottom panel: laboratory background.}
\end{figure}

\begin{figure}[t]
 \centering
  \includegraphics[angle=-90,width=1.0\columnwidth]{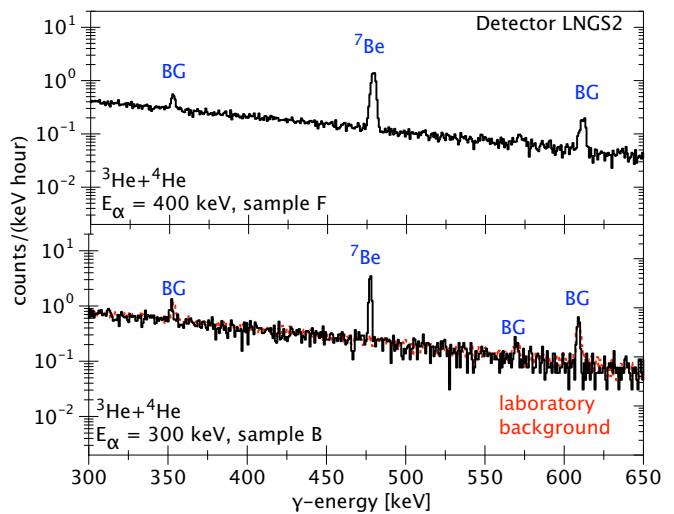}
 \caption{\label{fig:spectrum-lngs2} 
(Color online) Offline $\gamma$-counting spectra, detector LNGS2. Solid black line: $^3$He gas bombarded at $E_\alpha$ = 400\,keV (sample F) and 300\,keV (sample B), respectively. Dotted red line, bottom panel: laboratory background.}
\end{figure}

After the irradiation, the catcher is dismounted and counted subsequently with two 120\,\% relative efficiency HPGe detectors called LNGS1  (Fig.~\ref{fig:spectrum}) and LNGS2 (Fig.~\ref{fig:spectrum-lngs2}), both properly shielded with copper and lead, in the LNGS underground counting facility \cite{Arpesella96-Apradiso}. Detector LNGS1 is additionally equipped with an anti-radon box, and its laboratory background is two orders of magnitude lower than with equivalent shielding overground \cite{Arpesella96-Apradiso}. 

The samples have been counted in close geometry, e.g. in the case of detector LNGS1 the distance between sample and detector endcap was 5\,mm. In order to obtain a precise efficiency calibration at this close distance, three homogeneous $^7$Be sources of 200--800\,Bq activity and 8\,mm active diameter were prepared with the $^7$Li(p,n)$^7$Be reaction. Thin layers of LiF ($\simeq$20\,$\mu$g/cm$^2$) evaporated onto Ta backings and protected by evaporated gold layers ($\simeq5\,\mu$g/cm$^2$) have been irradiated by 2.5\,MeV protons from the ATOMKI Van de Graaff accelerator. The absolute activity of the calibration sources was subsequently determined in far geometry with two HPGe detectors at ATOMKI and with one HPGe detector, called LNGS3, at LNGS. The absolute efficiency of each of these three detectors has been determined using a set of commercial $\gamma$-ray calibration sources. The three source kits used for calibrating the detectors were mutually independent. All three  measurements gave consistent results, and the activities of the $^7$Be calibration sources have been determined with a final uncertainty of 1.8\,\%. 

The three $^7$Be calibration sources were then used to calibrate detectors LNGS1 and LNGS2 at close geometry. Owing to the relatively low activities of the calibration sources, random coincidence summing effect and deadtime correction were negligible. In the case of detector LNGS2 which has a horizontal geometry, calibration sources and samples were placed horizontally in front of the detector, and not vertically in top of it as in the case of detector LNGS1. The impact of statistical sub-millimeter variations in the distance between source/sample and detector endcap resulting from the horizontal geometry of detector LNGS2 has been evaluated by moving the calibrated sources. An additional uncertainty of 1.2\,\% resulting from this effect is included in the final statistical uncertainties given for detector LNGS2.

The $^7$Be distribution in the catchers has been calculated from the $^7$Be emission angle and straggling, and GEANT4 \cite{Agostinelli03-NIMA} simulations give 0.8$\pm$0.4\,\% to 1.5$\pm$0.4\,\% correction for the $\gamma$-ray efficiency because of the tail of the distribution at high radii.

\begin{figure}[b]
 \centering
  \includegraphics[angle=-90,width=\columnwidth]{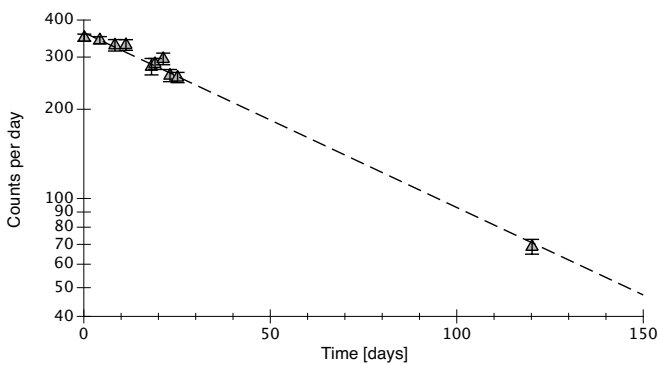}
 \caption{\label{fig:halflife-capA350} Counting rate of sample A on detector LNGS2 as a function of time. The dashed line is an exponential fit to the data.
 }
\end{figure}

Without aiming for high precision, the half life of $^7$Be in the Cu host material has been determined from the decay curve measured on the present, weak activated samples (Fig.~\ref{fig:halflife-capA350}). The weighted average of all measured samples gives a half life of 52.2\,$\pm$\,1.5~days, compatible with 53.22\,$\pm$\,0.06~days from a recent compilation  \cite{Tilley02-NPA}. The value from Ref. \cite{Tilley02-NPA} has been used for the data analysis.

The activity values referring to the end of the irradiations measured with the two HPGe detectors are in all cases in good agreement (Table~\ref{tab:activities}). For each sample, the weighted average of the activity values has been used for the data analysis.

\begin{table}[b]
\caption{\label{tab:activities} Counting times $\tau$ and activities measured for the different catchers with the two HPGe detectors. Uncertainties are purely statistical, in the case of detector LNGS2 also including the 1.2\,\% repositioning uncertainty discussed in the text.}
\begin{ruledtabular}
\setlength{\extrarowheight}{0.1cm}
\begin{tabular}{lrrclrrclrcl}
       & \multicolumn{4}{c}{LNGS1} & \multicolumn{4}{c}{LNGS2} & \multicolumn{3}{c}{Adopted}\\ \cline{2-5} \cline{6-9}
Sample & \multicolumn{1}{c}{$\tau$} & \multicolumn{3}{c}{Activity}  &  \multicolumn{1}{c}{$\tau$} & \multicolumn{3}{c}{Activity}   & \multicolumn{3}{c}{Activity} \\[-1mm]
 & [days] &  \multicolumn{3}{c}{[mBq]}    &   [days]            & \multicolumn{3}{c}{[mBq]} & \multicolumn{3}{c}{[mBq]} \\ \hline
D & 16 & 25.3 & $\pm$ & 1.3 & - & \multicolumn{3}{c}{-} & 25.3 & $\pm$ & 1.3 \\
B & 12 & 208 & $\pm$ & 6 & 21 & 203 & $\pm$ & 6 & 205 & $\pm$ & 4 \\
A & 6 & 472 & $\pm$ & 14 & 22 & 495 & $\pm$ & 11 & 486 & $\pm$ & 9 \\
F & 10 & 319 & $\pm$ & 11 & 11 & 310 & $\pm$ & 8 & 313 & $\pm$ & 6 \\
E & 16 & \multicolumn{3}{c}{$<$0.21 (2$\sigma$)} & - & \multicolumn{3}{c}{-} & 
\multicolumn{3}{c}{$<$0.21 (2$\sigma$)}
\end{tabular}
\end{ruledtabular}
\end{table}


\begin{figure}[t]
 \centering
  \includegraphics[angle=-90,width=1.0\columnwidth]{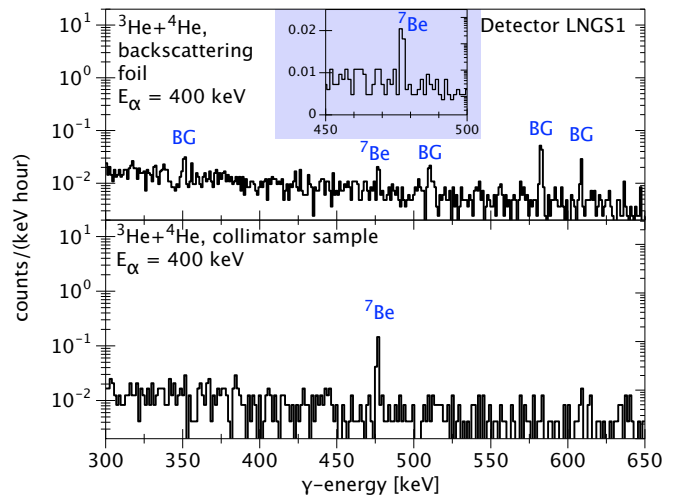}
 \caption{\label{fig:lngs1-collim+foils} 
(Color online) Offline $\gamma$-counting spectra, detector LNGS1. Collimator and backscattering foil counted in order to study $^7$Be losses. The peaks marked with BG in the upper panel are background lines from impurities in the Al foil. The inset in the upper panel has a linear ordinate.}
\end{figure}

\subsection{Parasitic $^7$Be production in the primary catcher}

Oxygen free high conductivity (OFHC) copper has been studied as a possible material for the primary catcher. This material has good heat conductivity (required for the beam calorimeter described above), sustains high $\alpha$-doses without blistering, and has a relatively low charge number in order to limit backscattering of $^7$Be nuclei. Since possible $^6$Li or $^{10}$B impurities in the catcher material can give rise to unwanted $^7$Be production through the $^6$Li(d,p)$^7$Be and $^{10}$B(p,$\alpha$)$^7$Be reactions induced by traces of $^2$DH$_2^+$ or even $^2$D$_2^+$ in the $^4$He$^+$ beam, this material was studied in detail prior to adopting it for the main experiment.

Samples from the material to be used for the catcher was irradiated with 700\,keV protons and deuterons at the ATOMKI Van de Graaff accelerator. After the irradiations, the $\gamma$-activity of the samples has been observed, showing no $\gamma$-peak from $^7$Be decay. Based on the known cross section of the $^6$Li(d,p)$^7$Be \cite{Szabo77-NPA,Hirst54-PM,Ruby79-NSE} and $^{10}$B(p,$\alpha$)$^7$Be \cite{Szabo82-conf,Burchmann50-PM} reactions, un upper limit of 3\,ppm has been determined for the concentration of both $^6$Li and $^{10}$B. By varying the settings of the LUNA2 analyzing magnet and taking the isotopic abundance of deuterium into account, an order-of-magnitude upper limit of 10$^{-7}$ d/$\alpha$ has been obtained. Combining this fraction with the above mentioned upper limits for $^6$Li and $^{10}$B contaminations and the $^6$Li(d,p)$^7$Be and $^{10}$B(p,$\alpha$)$^7$Be cross sections at lower energy, the induced $^7$Be activity from parasitic reactions is shown to be six orders of magnitude less than the activity from the $^3$He($\alpha,\gamma$)$^7$Be reaction expected using the $^3$He($\alpha,\gamma$)$^7$Be cross section from Ref. \cite{Kajino86-NPA}.

\begin{table*}[t!]
\caption{\label{tab:samples} Experimental (exp) and calculated (calc) corrections for irradiation and collection for each run. Effective energy and cross section are given in the final two columns. Only the statistical uncertainty is given. See Table \ref{tab:results} for the adopted systematic uncertainty.}
\begin{ruledtabular}
\setlength{\extrarowheight}{0.1cm}
\begin{tabular}{lrccccccc}
 & & &  \multicolumn{4}{c}{Corrections} \\ \cline{4-7}
Run & $E_\alpha$  & Target & Beam heating & Contaminant & Backscattering & Collection & $E^{\rm eff}$ & $\sigma(E^{\rm eff})$\\[-1mm]
 & [keV] & gas & & gases & & losses & [keV] & [10$^{-9}$\,barn] \\ \hline
D & 249.8 & $^3$He & 2.9\,\% & 0.5\,\%$_{\rm exp}$ & 2.9\,\%$_{\rm calc}$ & 2.4\,\%$_{\rm calc}$ & 105.6 & 0.567$\pm$0.029 \\
B & 298.8 & $^3$He & 4.9\,\% & 0.3\,\%$_{\rm exp}$ & 2.2\,\%$_{\rm calc}$ & 2.3\,\%$_{\rm calc}$ & 126.5 & 1.87$\pm$0.04\\ 
A & 348.4 & $^3$He & 5.4\,\% & 0.3\,\%$_{\rm exp}$ & 1.8\,\%$_{\rm calc}$ & 2.2\,\%$_{\rm calc}$ & 147.7 & 4.61$\pm$0.07 \\
F & 398.2 & $^3$He & 5.7\,\% & 1.0\,\%$_{\rm calc}$  & 1.3\,\%$_{\rm exp}$ & 2.6\,\%$_{\rm exp}$ & 168.9 & 9.35$\pm$0.19
\end{tabular}
\end{ruledtabular}
\end{table*}

The only excess activity detected on the samples irradiated at ATOMKI was $^{24}$Na produced by the $^{23}$Na(d,p)$^{24}$Na reaction. However, the half life of $^{24}$Na (15\,h) is short compared with that of $^7$Be, so in the main experiment the offline $\gamma$-counting spectra taken immediately after the irradiation was concluded were compared with spectra taken several days later, when any possible $^{24}$Na traces have decayed out. Thus, Compton background from the 2.754\,MeV $\gamma$-ray following the decay of $^{24}$Na has been ruled out as a significant contributor of background in the offline $\gamma$-counting.

Based on these considerations, OFHC copper was finally selected as material for the primary catcher. In order to rule out not only $^6$Li(d,p)$^7$Be, $^{10}$B(p,$\alpha$)$^7$Be, and $^{23}$Na(d,p)$^{24}$Na, but any possible source of parasitic $^7$Be, during the main experiment for one catcher the enriched $^3$He target gas was replaced with 0.7\,mbar $^{4}$He. This catcher was then bombarded at the highest available energy of
$E_\alpha$ = 400\,keV. Despite the high applied dose of 104\,C, in
16 days counting time no $^7$Be has been detected
(Fig.~\ref{fig:spectrum}, top panel, and Table~\ref{tab:activities}), establishing a 2$\sigma$ upper
limit of 0.1\% for parasitic $^7$Be.

\subsection{$^7$Be losses}

$^7$Be losses by backscattering from the primary
catcher and by incomplete collection were studied experimentally at
$E_\alpha$ = 400\,keV and with Monte Carlo simulations at 250, 300, 350
and 400\,keV. For the backscattering study, parts of the inner
surface of the chamber were covered by aluminum foil functioning as
secondary catcher (Fig.~\ref{fig:setup}\,e), and the foil sample was subsequently counted on detector LNGS1 (Fig.~\ref{fig:lngs1-collim+foils}, upper panel). It was found that
1.3\,$\pm$\,0.5\,\% of the created $^7$Be is lost due to
backscattering, consistent with 1.5\,\% obtained in a GEANT4
\cite{Agostinelli03-NIMA} simulation using a SRIM-like multiple
scattering process \cite{Mendenhall05-NIMB}. At lower energies, the
simulation result of up to 2.9\,\% was used as backscattering correction (Table~\ref{tab:samples}, column 7), with an adopted uncertainty of 0.5\,\%.

Incomplete $^7$Be collection occurs since 3.5\,\% of the total
$^3$He target thickness are in the connecting pipe, and a part of
the $^7$Be created there does not reach the primary catcher but is
instead implanted into the 7\,mm collimator
(Fig.~\ref{fig:setup}\,c). At $E_\alpha$ = 400\,keV, a modified
collimator functioning as secondary catcher was used and counted on detector LNGS1 (Fig.~\ref{fig:lngs1-collim+foils}, lower panel).
A 2.6\,$\pm$\,0.4\,\% effect was observed, consistent with a
simulation (2.1$\pm$0.4\,\%). For $E_\alpha$ = 250--350\,keV,
incomplete $^7$Be collection was corrected for based on the
simulation (up to 2.4\,\% correction, adopted uncertainty 0.5\,\%).

Sputtering losses of $^7$Be by the $^4$He beam were simulated
\cite{SRIM03-26}, showing that for the present beam energies
sputtering is 10$^{4}$ times less likely than transporting the
$^7$Be even deeper into the catcher, so it has been neglected.

All Monte Carlo calculations mentioned in sections II.D--II.F have been carried on until a statistical uncertainty of 0.2\,\% or better was reached, negligible compared to the systematic uncertainties discussed in the appropriate section.

\begin{table}[t]
\caption{\label{tab:uncert} Systematic uncertainties in the $^3$He($\alpha,\gamma$)$^7$Be astrophysical S factor. The uncertainty resulting from the effective energy affects only the S factor result, not the cross section.}

\begin{ruledtabular}
\setlength{\extrarowheight}{0.1cm}
\begin{tabular}{lc}
\multicolumn{1}{c}{Source} & Uncertainty\\
\hline
$^7$Be counting efficiency & 1.8\,\%\\
Beam intensity & 1.5\,\%\\
Beam heating effect  & 1.3\,\%\\
Effective energy & 0.5--1.1\,\%\\
Target pressure and temperature without beam & 0.8\,\%\\
Incomplete $^7$Be collection & 0.5\,\%\\
$^7$Be backscattering & 0.5\,\%\\
$^7$Be distribution in catcher & 0.4\,\%\\
478\,keV $\gamma$-ray branching \cite{Tilley02-NPA} & 0.4\,\%\\
$^7$Be half-life \cite{Tilley02-NPA} & 0.1\,\%\\
N$_2$ contamination in target gas & 0.1\,\%\\
Parasitic $^7$Be production & 0.1\,\%\\
\hline
Total: & 3.0--3.1\,\%\\
\end{tabular}
\end{ruledtabular}
\end{table}

\section{Results}

The effective center of mass energy $E^{\rm eff}$ has been calculated assuming a constant S factor over the target length \cite{Rolfs88}. The uncertainties of 0.3\,keV in $E_\alpha$ \cite{Formicola03-NIMA} and of 4.4\,\% in the energy loss \cite{SRIM03-26} lead to 0.16\,keV uncertainty in $E^{\rm eff}$ and thus contribute 0.5\,\% (at $E^{\rm eff}$ = 169\,keV) to 1.1\,\% (at $E^{\rm eff}$ = 106\,keV) to the S factor uncertainty. 

The effective energy and cross section results for each sample are shown in the last two columns of Table~\ref{tab:samples}. The systematic uncertainties are summarized in
Table~\ref{tab:uncert}, giving a total value of 3\,\%. For the
present low energies an electron screening enhancement factor $f$
\cite{Assenbaum87-ZPA} of up to 1.016 has been calculated in the
adiabatic limit, but not corrected for (Table~\ref{tab:results}).

\begin{figure}[t!]
 \centering
  \includegraphics[angle=270,width=\columnwidth]{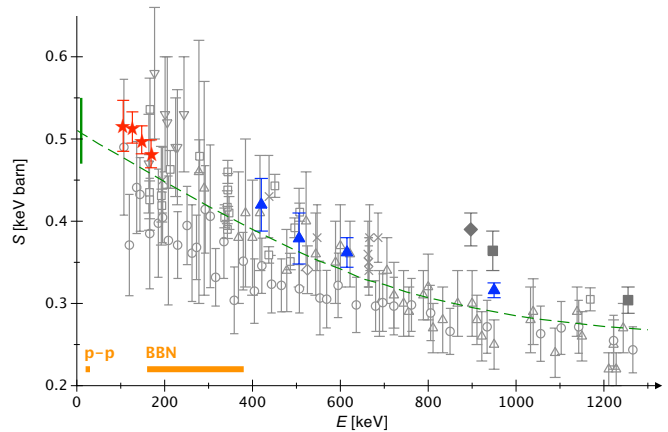}
 \caption{\label{fig:sfactor} (Color online) Astrophysical S factor for $^3$He($\alpha,\gamma$)$^7$Be.
Activation data: filled squares \cite{Osborne82-PRL}, filled diamonds \cite{Robertson83-PRC}, filled triangles \cite{NaraSingh04-PRL}, stars (present work).
Prompt-$\gamma$ data: triangles \cite{Parker63-PR}, inverted triangles \cite{Nagatani69-NPA}, circles \cite{Kraewinkel82-ZPA} (renormalized by a factor 1.4 \cite{Hilgemeier88-ZPA}), squares \cite{Osborne82-PRL}, diamonds \cite{Alexander84-NPA}, crosses \cite{Hilgemeier88-ZPA}.
Dashed line: previously adopted R-matrix fit \cite{Descouvemont04-ADNDT}. Horizontal bars: energies relevant for solar p--p hydrogen burning and for big bang nucleosynthesis.
 }
\end{figure}

The present data (Table~\ref{tab:results} and Fig.~\ref{fig:sfactor}) touch the energy range relevant to big-bang $^7$Li production. Their uncertainty of at most 6\,\% (systematic and statistical combined in quadrature) is
comparable to or lower than previous activation studies at high
energy and lower than prompt-$\gamma$ studies at comparable energy.

\begin{table}
\caption{\label{tab:results} Cross section and S factor results,
relative uncertainties, and electron screening
\cite{Assenbaum87-ZPA} enhancement factors $f$.}
\begin{ruledtabular}
\setlength{\extrarowheight}{0.1cm}
\begin{tabular}{rlccccc}
$E^{\rm eff}$ & $\sigma(E^{\rm eff})$ & $S(E^{\rm eff})$ & \multicolumn{2}{c}{$\Delta$$S$/$S$} & $f$ \\
\ [keV] & [10$^{-9}$ barn] & [keV barn] & stat. & syst. & \\
\hline
105.6 & 0.567 & 0.516 & 5.2\,\% & 3.1\,\% & 1.016 \\
126.5 & 1.87\footnotemark[1] & 0.514 & 2.0\,\% & 3.0\,\% & 1.012 \\
147.7 & 4.61\footnotemark[1] & 0.499 & 1.7\,\% & 3.0\,\%\footnotemark[2] & 1.009 \\
168.9 & 9.35\footnotemark[1] & 0.482 & 2.0\,\% & 3.0\,\%\footnotemark[2] & 1.008 
\footnotetext[1]{Cross section previously published in abbreviated form \cite{Bemmerer06-PRL}.}%
\footnotetext[2]{Systematic uncertainty 0.1\,\% higher than the one given in Ref. \cite{Bemmerer06-PRL}. The conclusions of Ref. \cite{Bemmerer06-PRL} are unaffected.}
\\
\end{tabular}
\end{ruledtabular}
\end{table}

\begin{table}[t!!]
\caption{\label{tab:Adelberger} Extrapolated S factor $S(0)$ from activation studies of $^3$He($\alpha,\gamma$)$^7$Be.}
\begin{ruledtabular}
\setlength{\extrarowheight}{0.1cm}
\begin{tabular}{llc}
& Ref. & $S(0)$ [keV barn]  \\ \hline
Osborne {\it et al.} & \cite{Osborne82-PRL} & 0.535$\pm$0.040 \\
Robertson {\it et al.} & \cite{Robertson83-PRC} & 0.63$\pm$0.04 \\
Volk {\it et al.} & \cite{Volk83-ZPA} & 0.56$\pm$0.03  \\
Nara Singh {\it et al.} & \cite{NaraSingh04-PRL} & 0.53$\pm$0.02 \\
present work & & 0.547$\pm$0.017  \\
\multicolumn{2}{l}{Weighted average, all activation studies} & 0.553$\pm$0.012 \\[2mm]
\multicolumn{2}{l}{Weighted average, all prompt-$\gamma$ studies \cite{Adelberger98-RMP}} & 0.507$\pm$0.016 \\
\end{tabular}
\end{ruledtabular}
\end{table}

\section{conclusion}

In order to obtain a recommended $S(0)$ value for the activation method, following Ref.~\cite{Adelberger98-RMP} it is instructive to list the extrapolated $S(0)$ values for the different activation studies together with their quoted uncertainty (table~\ref{tab:Adelberger}). For the present data, adopting the curve shape from Ref. \cite{Descouvemont04-ADNDT} an extrapolated $S(0)$ = 0.547$\pm$0.017 keV barn is obtained. The weighted average of all activation studies, including the present work, is found to be 0.553$\pm$0.012 keV barn, significantly higher than the weighted average of all prompt-$\gamma$ studies, 0.507$\pm$0.016 keV barn \cite{Adelberger98-RMP}.

With the addition of the new data, the systematic difference in normalization between prompt-$\gamma$ and activation studies of $^3$He($\alpha,\gamma$)$^7$Be is now smaller than in Ref.~\cite{Adelberger98-RMP}. However, it is still significant and much larger than the uncertainty required to match, e.g., the 3.5\,\% precision of the solar $^8$B neutrino data \cite{SuperKamiokande06-PRD}. 

In conclusion, prompt-$\gamma$ experiments with precision comparable to the present activation data are called for in order to verify the normalization of the prompt-$\gamma$ data.

\begin{acknowledgments}
This work was supported by INFN and in part by the European Union (TARI RII-CT-2004-506222), the Hungarian Scientific Research Fund (T42733 and T49245), and the German Federal Ministry of Education and Research (05CL1PC1-1).
\end{acknowledgments}

\end{document}